\newcommand{\fracj}[2]{{\textstyle{#1\over#2}}}
\newcommand{\rd}{\,{\rm d}}
\newcommand{\K}{\,{\rm K}}
\newcommand{\p}{\partial}
\newcommand{\s}{\,{\rm s}}
\newcommand{\erg}{\,{\rm erg}}
\newcommand{\kpc}{\,{\rm kpc}}
\newcommand{\kms}{\,{\rm km\,s}^{-1}}
\newcommand{\Myr}{\,{\rm Myr}}
\def\[{\begin{equation}}
\def\]{\end{equation}}
\documentclass{cflowproc}

\usepackage{natbib}

\begin{document}


\shortauthors{BINNEY}     
\shorttitle{COOLING FLOWS OR HEATING FLOWS?} 

\title{Cooling Flows or Heating Flows?}   

\author{James Binney   
}

\affil{}{Oxford University}   


\begin{abstract}
It is now clear that AGN heat cooling flows, largely by driving winds. The
winds may contain a relativistic component that generates powerful
synchrotron radiation, but it is not clear that all winds do so.  The
spatial and temporal stability of the AGN/cooling flow interaction are
discussed.  Collimation of the winds probably provides spatial stability.
Temporal stability may be possible only for black holes with masses above a
critical value. Both the failure of cooling flows to have adiabatic cores
and the existence of X-ray cavities confirm the importance of collimated
outflows. I quantify the scale of the convective flow that the AGN Hydra
would need to drive if it balanced radiative inward flow by outward flow
parallel to the jets. At least in Virgo any such flow must be confined to
$r\la20\kpc$.  Hydrodynamical simulations suggest that AGN outbursts cannot
last longer than $\sim25\Myr$. Data for four clusters with well studied
X-ray cavities suggests that heating associated with cavity formation
approximately balances radiative cooling. The role of cosmic infall and the
mechanism of filament formation are briefly touched on.
\end{abstract}


\section{Introduction}

How important are AGN for the dynamics of cooling flows? Are cooling flows
in approximate steady states, or do they evolve significantly over a Hubble
time? The debate of these issues is more than a decade old, but the recent
spectacular increase in the quality of the observational data has revived it
and stimulated a new generation of simulations of AGN/cooling-flow
interaction. Here I review these developments. In Section 2 I summarize the
observational situation. Section 3 discusses the physics of AGN/cooling-flow
interactions; each component affects the other and it is important to assess
whether these interactions are stabilizing or otherwise.  Section 4 gives
preliminary conclusions from a programme of large numerical simulations of
jet-induced heating that we are carrying out in Oxford.  Section 5 examines
the balance between heating and cooling from an empirical point of view and
concludes that, when averaged over $\sim100\Myr$, heating and cooling {\it
are\/} in balance.  Section 6 sums up and briefly discusses cosmic infall
and filament formation.

\section{The story so far}

Data from the Chandra and XMM-Newton missions have convinced pretty much
everyone that the intergalactic medium in clusters is not a
multi-temperature medium: although the quality of the fit to the spectrum of
an annulus can often be improved by including radiation from plasma at two
temperatures rather than one, the temperature of the cooler component never
differs by more than a factor of a few from that of the higher-temperature
component, and the emission measure of the hotter component always dominates
by more than an order of magnitude (Sanders this meeting).  Moreover, the
imaging data show significant inhomogeneities within annuli that require
two-temperature fits, and it seems likely that the second temperature
component merely compensates for temperature gradients within each annulus.
These results falsify the picture of distributed mass dropout that dominated
cooling-flow studies for nearly two decades.

Within a few kpc of the cluster centre, cool gas {\it is\/} detected. Gas at
$\la100\K$ is detected in the rotation lines of CO
\citep{Lazareff89,MirabelSK,Reuter93,Edge01}. Gas at a few thousand Kelvin
is detected in the rotation-vibration transitions of H$_2$
\citep{Donahue_etal00,EdgeEtal02}.  Gas at a few times $10^4\K$ is detected
through optical emission lines
\citep{Lynds70,CowieEtal83,Heckman89,Conselice01}. Gas at $\sim10^6\K$ is
detected in soft X-rays \citep{FabianCrawford03}. The spatial distribution
of the observed emission lines and soft X-rays is consistent with the cold
gas being organized into filaments, within which the temperature increases
continuously from the centre outwards. The filaments have complex velocity
fields and it seems very unlikely that they are in dynamical equilibrium
within the cluster potential. The mass of gas at say a tenth of the
cluster's virial temperature is at least an order of magnitude smaller than
the mass of such gas that was predicted to accumulate over the cluster
lifetime by the distributed dropout model. Moreover, is much more centrally
concentrated than it was predicted to be. 

In parallel with destroying the concept of distributed mass dropout, the new
data provide clear evidence that active galactic nuclei are pumping energy
into cooling flows. Specifically, the new data show that cavities in the
X-ray emitting gas, like that discovered by \cite{BohringerEtal93} in the
Perseus cluster, are common
\citep{FabianSET,BlantonEtal01,McNamaraEtal01,NulsenDMJFW,HeinzCRB}. The low
X-ray emissivity within these cavities must arise because the pressure
within them is dominated either by very hot thermal plasma or by
relativistic particles and their associated magnetic fields, or by both.
Radio emission from many cavities shows that these objects are regions of
enhanced energy density from ultra-relativistic particles and fields, which
establishes a clear connection between the cavities and activity by AGN
within the cluster.

The title of this conference, `The Riddle of Cooling Flows', reflects a
widespread feeling that these observational results are unexpected and
perplexing. I do not share this feeling because I have been arguing for over
a decade (i) that distributed mass dropout makes no sense physically and
(ii) that cooling can occur only at the centre, where it will generate
outbursts by the AGN, so (iii) there is a fundamental symbiosis between
cooling flows and AGN. I say this not merely to brag (though I do) but to
make the point that  far from being puzzling, the data from Chandra
and Newton-XMM are very much in line with expectation if one applies sound
physical principles to systems in which there is gravitationally trapped
cooling gas.

A decade ago my student Gavin Tabor worked out
detailed models of cooling flows in elliptical galaxies rather than clusters
of galaxies because I felt surer of the physics in smaller systems: the
timescales are very short in elliptical galaxies and the current impact of
system formation and cosmic infall can be much more securely neglected.
Unfortunately, the quality of the data for elliptical-galaxy cooling flows
is still very poor [the exceptions being Cen A \citep{Kraftetal03} and NGC
4636 \citep{Kahn02}] but I
have no doubt that a similar picture applies in ellipticals, and strongly
suspect that the well known correlations between nuclear black hole masses and global
properties of bulges involves the dynamics of galaxy-sized cooling flows in
an essential way \citep{binney_nocool}.

\section{AGN as heaters}

We now know for certain that massive black holes lurk at the centres of
cooling flows. They will accrete ambient gas. The natural estimate of the
rate at which a black hole accretes from an atmosphere of given pressure $P$
and sound speed $c_s$ is the \cite{Bondi} accretion rate
 \[\label{binney:BondiM}
\dot M=4\pi\left({GM\over c_s^2}\right)^2\rho c_s
=4\pi G^2M^2{\gamma P\over c_s^5}.
\]
 This accretion rate is extremely sensitive to the sound speed of the
atmosphere near the black hole.  This sensitivity makes it plausible that
the black hole is a thermostatically controlled heater: when radiative
cooling lowers $c_s$, the accretion rate rapidly rises and releases
accretion energy that reheats the atmosphere. Notice that the accretion rate
depends on $c_s$ even more sensitively than the fifth power manifest in
(\ref{binney:BondiM}) because $P$ will rise slightly as $c_s$ declines.
However, the dependence of $\dot M$ on $c_s$ through $P$ is weak, and below
I shall ignore it.

\subsection{Stability}

Suppose a cooling flow is in a state in which heating balances cooling at
each radius and consider the stability of this state in both a spatial and a
temporal sense. The question with regard to spatial stability is this.
Suppose the heating weakens for some reason, then the net cooling that
ensues will make the density profile steeper than it was. Will the system
recover from this excursion when the heating rate later increases? The
answer to this question is `no' if the AGN merely returns to its former
power: that power balanced the unmodified cooling rate, so it will be less
than the enhanced current cooling rate.  But if the AGN returns to work with
renewed vigour after its rest, refreshed perhaps by the enhanced Bondi
accretion rate, then I think it is plausible that the system can recover its
old density profile provided it heats the ambient medium with jets. For in
this case the spatial distribution of the injected energy depends on the
density profile of the cooling gas. Enhanced density at small radii leads to
the jets disrupting closer to the AGN, and thus increases the fraction of
the AGN's power that is dissipated at small radii. If the increased
concentration of the jets' energy deposition is large enough, gas near the
AGN will expand faster than gas further out, and the density profile will
flatten to its former shape. Clearly, this mechanism needs to be quantified
by hydrodynamical simulations.

\begin{figure}
\plotone{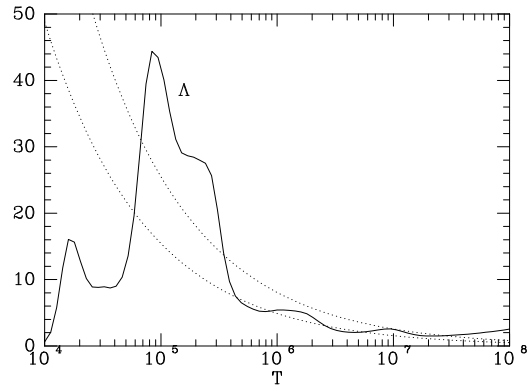}
\figcaption{Full curve: the cooling coefficient $\Lambda(T)$ for a plasma
with $0.4$ of solar abundances. Dashed curves: curves proportional to
$T^{-1/2}$ that intersect the full curve at $T=3.5$ and $4.5\times10^5\K$.
\label{binney:fig1}}
\end{figure}

To investigate the temporal stability of heated cooling flows, consider the
differential equation 
 \[\label{binney:Tdot}
{\rd T\over\rd t}={1\over T^2}\left({A\over\surd T}-\Lambda(T)\right).
\]
 This equation describes the temperature of a system that is heated at a
rate that scales $\sim T^{-5/2}$ [cf.\ eq.~(\ref{binney:BondiM})] and is
cooled at a rate that scales as $\sim T^{-2}\Lambda(T)$, as does a plasma
that is confined by a constant pressure.  In Figure \ref{binney:fig1} the
two terms inside the big bracket on the right of equation
(\ref{binney:Tdot}) are plotted, the dotted curves showing the first term
for two values of the constant $A$. Thermal equilibrium is possible when the
dotted and full curves cross; for the equilibrium to be stable the full
curve has to have the larger slope at the point of intersection. We see that
for $A$ less than some maximum value, stable equilibria exist at
temperatures less than that of the peak in the cooling curve at $83\,000\K$.
For any value of $A$ at least one point of stable equilibrium exists at a
higher temperature. In the case of the upper dotted curve, a unique
high-temperature equilibrium lies at $2.7\times10^7\K$.  For the lower dotted
curve there are stable equilibria at $10^6\K$ and $5\times10^6\K$.  The
stability of an equilibrium depends on the difference in the slopes of the
intersecting curves, so  the high-temperature equilibria are
only weakly stable in a linear sense, and, in the case of the lower dotted
curve, can be destabilized by quite modest variations in the parameter $A$.

The low- and high-temperature equilibria differ strongly in the system's
accretion rate and luminosity: the low-temperature equilibrium has the larger
luminosity by a factor $f^{5/2}$, where $f$ is the ratio of the
temperatures; in the case of the upper dotted curve in Figure \ref{binney:fig1} this factor is
$3.4\times10^6$.

Equation (\ref{binney:Tdot}) is only a toy equation that gives some insight
into a complex dynamical system that is governed by non-linear hydrodynamic
equations. The temperature that appears in the equation should be
interpreted as the temperature of the accreting plasma at the point inside
the black hole's sphere of influence where the Kepler speed equals the sound
speed. This temperature will not be significantly lower than that, $T_{\rm
min}$, associated with the Kepler speed at the edge of the black hole's
sphere of influence. Only in very small galaxies will $T_{\rm min}$ be less
than $10^5\K$. Hence real accreting black holes may not possess the
low-temperature equilibrium that the toy equation (\ref{binney:Tdot})
predicts. What the equation does correctly predict for systems with small
values of $A$ is their {\it instability} at all temperatures in the range
$0.8$ to $5\times10^5\K$ and roughly half of the temperatures from there up to
$2\times10^7\K$.  This wide-ranging instability may well lead to runaway
growth of the accretion rate and luminosity.

The coefficient $A$ that determines what equilibrium points exist reflects
the relative importance of heating and cooling. An increase in the mass of
the accreting black hole within a given cooling flow will be reflected in an
increase in the best-fitting value of $A$, and may move $A$ to a value at
which only a stable high-temperature equilibrium is possible. Hence our toy
equation suggests that black holes above a critical mass (that depends on
the parameters of the cooling flow) have  a single moderately stable state of
low luminosity, while lower-mass black holes have one or more low-luminosity
states of marginal stability and are otherwise liable to experience
runaway growth of their luminosity.

\subsection{Output channels from AGN}

We still do not have a clear picture of what happens when a given mass of
gas falls onto  a massive black
hole. Equation (\ref{binney:BondiM}) is a plausible estimate of the rate at
which gas falls from the centre of a cooling flow into the force-field of
the central black hole. But how much energy is released in consequence, and
in what form does it emerge?

The efficiency $\epsilon=\dot E/(\dot Mc^2)$ with which accretion by a black hole releases
energy is controversial.  Comparison between the space density of massive
black holes and the integrated luminosity density of quasars implies
$\epsilon>10\%$ in luminous quasars \citep{YuT}. Controversy rages as to
whether a similar efficiency applies to accreting systems that are faint in
the optical and UV. It has long been recognized that implausibly large
luminosities are derived for many massive black holes if the rate
(\ref{binney:BondiM}) is multiplied by an efficiency anywhere near as high
as $10\%$ \citep{FabianCanizares88}. One school of thought argues that around these black
holes, which include Sgr A$^*$ at the Galactic centre and the black holes at
the centres of all nearby cooling flows, electrons fail to heat and the ions
carry the accretion energy over the event horizon before the electrons can
radiate it.  To my mind this picture, like distributed mass dropout, is
physical nonsense \citep{binneyADAF}. A much more plausible explanation for
the low luminosities of many black holes is that offered by
\cite{BlandfordBegelman99}: a wind from the surface of the accretion disk
carries off much of the energy and angular momentum that is released by
accretion. Moreover, the wind carries away most of the mass that falls
within the sphere of influence of the black hole at the rate given by
(\ref{binney:BondiM}). The upshot is rather a small rate of mass flow over
the event horizon, and the release of nearly all the associated accretion
energy as a sub-relativistic wind. 

We are familiar with synchrotron-emitting jets emanating from the nuclei of
the central galaxies of cooling flows. It is natural to assume that the
Blandford-Begelman wind is collimated on parsec scales and to identify this
with the observed jet. But this association is dangerous!  The jets from M87
and similar galaxies have significant bulk Lorentz factors $\gamma$, while a
wind from the accretion disk is expected to be non-relativistic.
Consequently the observed jets probably arise through vacuum breakdown in
the ergosphere of a rotating black hole \citep{BlandfordZ} and are not directly
connected to the subrelativistic wind from the disk.  The observed jets are
conspicuous because shocks in them readily generate synchrotron-emitting
electrons, but they may be ephemeral side-shows. In particular, they may
flicker on and off on the short timescale associated with the ergosphere,
and we should expect to see systems in which the wind is present, but there
is no relativistic jet. Sgr A$^*$ may be just such a system.  The connection
between the mechanical luminosity of the sub-relativistic wind and the
synchrotron luminosity of the source may be weak or non-existent. Jet
luminosities derived from observations of relativistic jets, such as that
for M87 by \cite{ReynoldsFCR} constitute lower limits on the total
mechanical luminosity of the AGN.

\subsection{Collimation}

We should keep an open mind about the extent to which the sub-relativistic
wind is collimated. The less well collimated it is, the more locally it will
couple to the thermal plasma of a cooling flow. Tabor first investigated the
case of local coupling that would apply if the wind were essentially
uncollimated \citep{TaborBinney93}. In this case the wind would shock near
the black hole to produce a hot cavity -- a similar cavity would be produced
in the situation envisaged by \cite{CiottiOstriker97,CiottiOstriker01}, in
which the flow is heated by inverse-Compton scattering of photons.  An
entropy inversion is created in which specific entropy decreases outwards.
The region in which this occurs is convectively unstable. As in an
early-type star, convection cells soon eliminate this gradient, and the
system settles to approximate hydrostatic equilibrium in a configuration
that has a constant-entropy core. Tabor \& Binney constructed models in
which this constant entropy core joined at some radius to a steady
cooling-driven inflow.

The new data have shown that  in cooling flows specific entropy is not
constant near the centre: in reality the mass $M(\sigma)$ with entropy index
$\sigma=P\rho^{-\gamma}$ less than some value fits the formula
 \[\label{binney:Msigma}
M\propto(\sigma-\sigma_0)^\epsilon
\]
 with $\sigma_0$ a constant and $\epsilon\sim1.5$
\citep{KaiserBinney03a,KaiserBinney03b}. This finding suggests that the
outflow from the AGN that heats a cooling flow is collimated into jets.  A
jet simultaneously heats the ambient medium at several different radii,
and thus offers an alternative to convection as a means of carrying heat
from the AGN out to the sites of radiative cooling.  The observed cavities
in the X-ray emitting gas provide direct evidence for collimation.

\subsection{FR I or FR II sources?}\label{binney:FRI}

The radiation of X-rays makes the cluster gas more centrally concentrated,
and if unchecked will produce a central cooling catastrophe in a typical
cooling flow within several hundred Myr. \cite{BinneyTabor95} conjectured
that at some stage the central AGN `winds back the clock' and restores the
cooling flow to the state it was in at an earlier time. A few powerful
outburst could set the clock back several Gyr each, or many weaker outbursts
could set the clock back tens of Myr each.

Fanaroff-Riley (FR) II radio galaxies have powerful jets that end in a hot
spot on the periphery of the source, while the weaker jets of FR I sources
break up or bend dramatically near the centre of the source.
FR II sources are thought to have kinetic luminosities $\sim100$ times the
cooling-flow luminosity and lifetimes $\sim100\Myr$, so each outburst
injects energy comparable to that radiated in X-rays over a Hubble time.
Consequently, they can be responsible for reheating cooling flows only if
they radically reduce the cooling flow's central density to the point at
which the central cooling time becomes almost as long as the Hubble time.
\cite{ReynoldsHeinzBegelman02} and \cite{BassonA} have simulated the impact
on intracluster gas of FR II sources and shown that most of the energy is
injected well beyond the cooling radius, so FR II sources do not  reduce the density
within the cooling radius  sufficiently.

Eilek (this meeting) notes that the great majority of sources at the centres
of cooling flows are of the FR I type. Hence they have relatively small
luminosities and jets that become unstable at a small fraction of the
cooling radius. In Virgo \cite{OwenEilekKassim00} have mapped the diffuse
synchrotron emission to unprecedentedly low surface-brightness levels, and
shown that the emission has a sharp edge at $\sim0.4r_{\rm cool}$. This
finding suggests that not only the current outburst, but all its
predecessors have dissipated their energy within this radius. Thus we have a
consistent picture in which regular weak outbursts by FR I sources replenish
the energy radiated in the innermost part of the cooling flow.  Our group in
Oxford is concentrating on simulating the impact of weak FR I sources
\citep{OmmaBBS}, and below I give preliminary conclusions from this work.

The existence and dynamics of FR II sources implies that the thermostat on a
massive black hole can fail: a rapid drop in the central sound speed $c_s$,
caused, for example, by a major accretion event, will provoke a powerful
nuclear outburst.  However, precisely the power of the jets will cause them
to burst out of the core of the cooling flow and thus strongly diminish
their ability to heat and expand the gas that is causing the outburst. Thus
while cooling flows may be stable to small perturbations, sufficiently large
perturbations may cause them to flare up dramatically.

\subsection{Local heating versus entrainment}

One can imaging two rather different routes by which an AGN might establish
a steady state. In the most straightforward route it would inject into each
radial shell heat that balanced the local cooling. In practice a major
contributor to the heating of the X-ray emitting plasma would be the
dissolution of  cavities.  Another possibility is that entrainment of
ambient gas would establish an outflow along the jet axis that balanced a
radiatively driven inflow elsewhere.  The synchrotron map of Virgo by
\cite{OwenEilekKassim00} cited above limits the applicability of this model
in the case of Virgo: since outflow appears not to extend beyond
$r\sim40\kpc$, the regime in which dissolution of cavities is unimportant
cannot be relevant outside $r\sim20\kpc$, although it might apply at small
radii. Heating by dissolving cavities {\it must\/} be important at radii of
order $30\kpc$. 

To understand how cavities dissolve, imagine pouring a
tanker full of olive oil into the ocean. The oil would quickly flow over the
ocean surface to form a huge sheet, only a few molecules thick. In a
similar manner the fluid inside a cavity spreads out azimuthally when it
reaches the level at which the ambient medium has a density similar to
itself. What makes the process more complex than the spreading of oil on
water is the absence of a sharp density discontinuity in the ambient medium
analogous to the water/air interface, and ongoing turbulent mixing of hot and
cold plasma as the cavity rises and dissolves.

\begin{figure}
\plotone{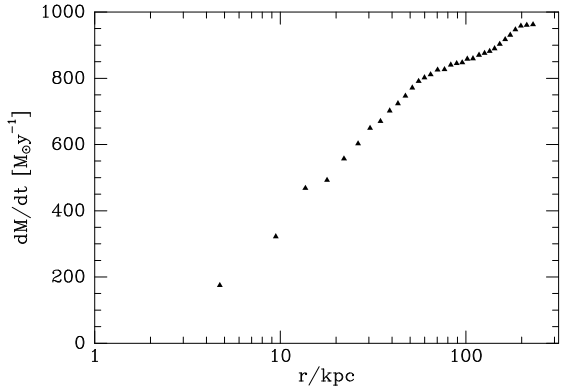}
\figcaption{The $\dot M$ profile for the Hydra cluster from the data of
\cite{DavidEtal00} and equations (\ref{binney:Mdot}) and
(\ref{binney:Ltos}). The metallicity is assumed to be $0.4Z_\odot$.
\label{binney:fig2}}
\end{figure}

To investigate the regime of balanced in- and out-flows that might apply far
inside the cooling radius, it is useful to calculate a different
$\dot M$ profile to one that is familiar from the discredited mass-dropout
picture.  Consider a body of gas, instantaneously in hydrostatic
equilibrium, in which the specific entropy density $s$ increases outwards.
In the absence of heating, the rate at which mass moves through specific
entropy $s$ is
\[\label{binney:Mdot}
 {\rd M\over\rd t}\bigg|_s={\p m\over\p s}{\rd s\over\rd t}.
\]
 I evaluate the first derivative on the right using (\ref{binney:Msigma}),
and the second derivative follows from the luminosity per unit mass
 \[\label{binney:Ltos}
L={\Lambda(T) n_e^2\over n_em_p}=T{\rd s\over\rd t}.
\] 
 Fig.~\ref{binney:fig2} shows the resulting $\dot M$ profile for Hydra.
This profile resembles the $\dot M$ profile of traditional mass-dropout
theory in increasing with radius, but there are important differences of
detail between the two profiles. First, interior to $r\sim50\kpc$ the
profile of Fig.~\ref{binney:fig2} shows that
 \[\label{binney:Mdotofr}
\dot M\propto\log{(r/r_0)}
\]
 with $r_0\sim2.5\kpc$, while a conventional $\dot M$ profile
has $\dot M\propto r$. Thus the mass flux rises more slowly with radius than
in the conventional picture.  Second, the profile in Fig.~\ref{binney:fig2}
reaches values of $\dot M$ that are about twice as large as those reached by
the traditional profile for this cluster \citep{DavidEtal00}. Both results
arise because we are calculating not the rate at which mass crosses a given
radial shell, but the rate at which it passes through a given specific
entropy. Since we have allowed for cooling but not heating, the temperature
and entropy at a given radius are decreasing, so the radius at which a given
value of the specific entropy is attained is moving outwards at the same
time that each physical shell of gas is moving inwards. Hence, we are
determining the mass flux of a flow with respect to a moving frame of
reference. The frame moves fastest near the centre, so that is where $\dot
M$ is most enhanced with respect to the traditional value.

If radiatively driven inflow is balanced by outflow along the jet, the mass
flux carried by the outflow would have to increase by $(\rd\dot M/\rd s)\rd
s$ between the radii at which the specific entropy of the inflowing gas
moved from $s$ to $s+\rd s$. If we assume that (\ref{binney:Mdotofr}) holds
and that the radial density profile can be approximated by a power law (in
practice $\rho\propto r^{-1.5}$ in a wide range of $r$), then one can show
that
 \[
{\rd\dot M\over\rd s}=\hbox{constant  for }\sigma\gg\sigma_0.
\]
 The numerical value of the mass-flow rate along the jet axis would be given
by Fig.~\ref{binney:fig2}. It would be very large even at the small radii
($r\la30\kpc$) at which this picture might apply.

\section{The Oxford simulations}

Advances in computer hard- and software finally make it feasible to run the
simulations of cooling-flow heating that I would have liked Tabor to run a
decade ago. Then we had to settle for spherically symmetric hydrodynamics,
which, as we stressed, inevitably excluded much of the key physics. We now
use ENZO \citep{BryanNorman97}, a  code which does hydrodynamics on adaptive Cartesian
grids.  As far as we are aware, ours is the only work on this problem that
uses three-dimensional adaptive grids to achieve kpc-scale resolution at the
centre of a box that encompasses the entire cluster.

For the reasons given in Section \ref{binney:FRI}, we concentrate on
low-power outbursts and seek to understand the dynamics of cavities that are
confined to within the cooling radius. Previous simulations aimed at this
problem \citep{ChurazovEtal01, QuilisEtal01, BruggenKaiser01, BruggenKCE02,
BruggenKaiser02} have simply deposited energy at some arbitrarily chosen
location. In our simulations \citep{OmmaBBS} we inject both energy and
momentum in such a way that jets form within the simulation, and these then
heat the ambient medium in two dynamically determined regions. In the
simulations of \cite{ReynoldsHeinzBegelman02} and \cite{BassonA}, the
regions of heating were also dynamically determined. However, these
simulations differ from ours in two respects: (i) their jets were $>100$
times more luminous than ours, and (ii) their jets were imposed through
inflow boundary conditions on a sphere that excluded the cluster centre from
the computational region. Our algorithm for
jet formation avoids the use of internal boundary conditions.

Our main conclusions from this ongoing work are the following

\begin{enumerate}

\item 
The duration of an AGN outburst, rather than its power, is what determines
the range of radii within which its energy is finally deposited.  Unless AGN
outbursts are of rather short duration ($\la30\Myr$), cavities move beyond
the cooling radius and deposit their energy further from the centre than is
required if heating is to offset cooling. 

\item
Entrainment of cool gas by the jet is an important process, as is turbulent
mixing of jet-heated and ambient gas.

\item
A turbulent vortex that contains a significant quantity of cool gas trails
each cavity. After the jet has switched off, the advance of the cavity slows
and the vortex overtakes it. When it is about twice as old as the duration
of the outburst, the cavity becomes an overdensity.

\item
The overdensity overshoots the radius in the cooling flow at which the
ambient medium has the same specific entropy. Then it falls back and excites
the strong internal gravity waves in the cooling flow that accompany its
azimuthal spread and disappearance.

\item
Interactions between AGN outbursts are very important. If an outburst follows
the last sooner than $\sim70\Myr$, the new jet is disrupted at small radii by
the turbulent wake of its predecessor and no new cavity forms. 

\item
Sheets of cold dense gas on the leading edges of cavities, similar to those
observed as X-ray bright rims, frequently form as a cavity is driven up
through cold gas that is rising in the wake of the previous cavity.

\end{enumerate}

\begin{table*}
\begin{center}
\caption{Parameters for five clusters with cavities.}
\begin{tabular}{lcccl}
System&$PV/10^{58}\erg$&$L_X/10^{43}\erg\s^{-1}$&$\tau/\Myr$&Reference\\
\hline
Hydra A&27&30&	88	&\cite{McNamaraEtal00,NulsenDMJFW}\\
A2052  &4 &3.2&	122	&\cite{BlantonEtal01}\\
Perseus&8 &27  &29		&\cite{FabianSET,AlenFJEN}\\
A2597  &3.1&3.8&79	&\cite{McNamaraEtal01}\\
A4059  &22&18&	119	&\cite{HuangS,HeinzCRB}\\
\hline
\end{tabular}
\end{center}
\end{table*}

\section{A statistical steady state?}

The traditional view is that cooling flows are in steady states, with
cooling balanced by mass-dropout, but a statistical steady state could be
achieved by a sequence of small outbursts by the AGN \citep{TaborBinney93}.
A radically alternative point of view is that cooling flows experience a
series of cooling catastrophies that provoke powerful AGN outbursts, which
rapidly inflate the core of the cooling flow to a significant extent,
setting the scene for a prolonged drift towards the next cooling catastrophe
\citep{BinneyTabor95,CiottiOstriker97,CiottiOstriker01,KaiserBinney03a}.
Simulations of the impact of FR II sources on cooling flows
\citep{ReynoldsHeinzBegelman02,BassonA} imply that powerful AGN outbursts of
the type we observe locally do not heat cluster cores sufficiently to ensure
a long interval before the next catastrophe. Less strongly collimated
outbursts may achieve longer intervals between catastrophes, but currently a
succession of weak outbursts seems the most likely possibility.  Moreover,
recent observations of X-ray cavities support this conclusion.

If we model the energy density within a cavity by that of a non-relativistic
gas, then the {\it minimum\/} work done by an AGN in blowing a cavity of
volume $V$ in a medium of pressure $P$ is $\fracj52PV$, of which three
fifths resides in the cavity.  If the fluid within the cavity is
relativistic, the minimum energy is $4PV$.  Since the inflation of cavities
is likely to be highly irreversible, especially in its early stages and from
the perspective of the ambient medium, the actual work done will be larger.
Below I shall rather conservatively assume that the work done is $3PV$.
\cite{ChurazovSFB} discuss the physics of cavity inflation in some detail.
As a cavity is blown, energy is carried away into the ambient medium by
hydrodynamical motions, the high-frequency tail of which will be recognized
as non-linear sound waves. Such waves may recently have been observed by
\cite{FabianEtal03} in X-ray images of the Perseus cluster. The bulk motions
associated with lower frequencies may manifest themselves in the peculiar
velocities of H$\alpha$ filaments \citep{Heckman89}.  Much of the energy acquired by
ambient gas as a cavity inflates will be used to lift gas up in the cluster
potential.  The cooling time of the gas will be increased but the gas may
not actually heat \citep{Alexander02}.

Values for the product $PV$, the X-ray luminosity of the cooling flow $L_X$
(mostly from the classic $\dot M$ value) and the characteristic time
$\tau=3PV/L_X$ are given in Table 1. We see that heating associated with
cavities can offset radiation from the cooling flow provided outbursts occur
every $30$ to $120\Myr$. Simulations show that cavities rise at of order the
sound speed $\sim1000\kms$, so they move
through $\sim50\kpc$ in $50\Myr$. Given the high frequency with which
cavities are found in the cores of cooling flows, it is clear that a new
pair must be created every $\la100\Myr$. Arguments from synchrotron ageing
have long implied that more powerful radio sources have lifetimes of this
order \citep{Pedlar90}, but here the conclusion is slightly different:
individual AGN outbursts might last a significantly shorter time that
$100\Myr$, but the interval {\it between\/} outbursts is of this order. 

The bottom line of this discussion it that, in a statistical sense, heating
by AGN may well balanced radiative cooling, as \cite{TaborBinney93}
originally conjectured. However, the issue cannot be regarded as closed because
as Burns (this meeting) and Thomas (this meeting) have pointed out,
cosmological simulations imply that clusters are accreting relatively cool gas. The power required to heat this infalling gas to
the virial temperature will significantly increase the power an AGN has to
provide if it is  to keep its cooling flow in a steady state. The big
uncertainty here is how much of the infalling gas reaches the clusters
radiatively cooling core. Gas that ends up outside the core can be largely ignored.

\section{Conclusions}

The Chandra and XMM-Newton missions have opened the way a clear understanding of
how cooling flows work. Distributed mass dropout has been decisively
rejected and the importance of heating by AGN established. The data seem to
show that there is approximate balance between AGN heating and radiative
cooling. Although the heating does have to be episodic, the time between
heating episodes is significantly shorter than even the central cooling time,
so the radial density profiles of systems do not change greatly around a
cooling/heating cycle. 

The favoured heating agent is collimated outflow from the AGN: a collimated
outflow can heat ambient material at several radii and values of specific
entropy simultaneously. Consequently there is no need for the system to
develop a constant-entropy convective core. It is probably also important
for the stability of a cooling flow that it is heated by jets: steepening of
the ambient density profile causes the jets to disrupt at smaller radii. The
resulting increase in the concentration of the heat source may well be
successful in counteracting the increase in the concentration of the cooling
rate that accompanies a steepening of the density profile. 

Much of the energy in the heating outflow will be contained in a
sub-relativistic wind.  The opening angle of this wind may not be small. At
its core there may or may not be relativistic jets. Observations of
synchrotron radiation will be largely sensitive to the relativistic jets and
their fall-out, so such observations may provide a very incomplete picture of
the overall AGN/cooling flow interaction.

Many questions remain open. Two of the most interesting, and possibly
connected, issues are the roles of cosmic infall and cool filaments. Have
the filaments formed through condensation of X-ray emitting gas, as has
traditionally been argued \citep{FabianN}? If this conjecture were correct,
and in view of the thermal stability of gravitationally stratified plasma,
it is surprising that the filaments seem so far from dynamical equilibrium:
one would expect condensation to occur onto a rotationally supported disk
rather than often radially directed filaments. Also, as \cite{SparksEtal89}
have stressed, one would not expect gas that had condensed from the X-ray
emitting plasma to have dust that is similar to Galactic dust. Perhaps
filaments reflect the cosmic infall of cool gas and gas-rich galaxies. Very
near the centre of the cluster this gas might take so long to evaporate that
it can be lighted up by internal star formation and then be observed (Nipoti
\& Binney in preparation).  In this case, the AGN would not normally be
accreting gas from filaments, but feeding directly on the X-ray emitting
gas. The AGN's luminosity would then be closely connected to the temperature
of the coolest X-ray emitting gas, and the efficient operation of
thermostatic feedback would be natural and fairly well modelled by equation
(\ref{binney:Tdot}).


\acknowledgements
I thank Henrik Omma for many valuable insights.
A grant from Merton College enabled me to attend this meeting.


\end{document}